\documentclass[pre,  twocolumn, showpacs,amssymb,superscriptaddress]{revtex4-1}

\usepackage{epsfig, color}
\usepackage{amsthm, amsmath, amsfonts, ae, psfrag}
\usepackage{braket}
\usepackage{bbm}
\usepackage[english]{babel}
\usepackage{epstopdf} 
\usepackage{listings}
\usepackage{placeins}	
\usepackage{verbatim} 

\usepackage[colorlinks,
pdfpagelabels,
pdfstartview = FitH,
bookmarksopen = true,
bookmarksnumbered = true,
linkcolor = blue,
plainpages = false,
hypertexnames = false,
citecolor = blue] {hyperref}

\begin{document}

\title{Bulk dynamics of Brownian hard disks: Dynamical density functional theory versus experiments on two-dimensional colloidal hard spheres}

\author{Daniel Stopper}
\email{daniel.stopper@uni-tuebingen.de}
\affiliation{Institute for Theoretical Physics, University of T\"{u}bingen, 72076 T\"{u}bingen, Germany}
\author{Alice L. Thorneywork}
\affiliation{Department of Chemistry, Physical and Theoretical Chemistry Laboratory, University of Oxford, South Parks Road, Oxford OX1 3QZ, United Kingdom}
\affiliation{Cavendish Laboratory, University of Cambridge, Cambridge CB3 0HE, United Kingdom}
\author{Roel P. A. Dullens}
\affiliation{Department of Chemistry, Physical and Theoretical Chemistry Laboratory, University of Oxford, South Parks Road, Oxford OX1 3QZ, United Kingdom}
\author{Roland Roth}
\affiliation{Institute for Theoretical Physics, University of T\"{u}bingen, 72076 T\"{u}bingen, Germany}

\date{\today}

\begin{abstract}

Using dynamical density functional theory (DDFT), we theoretically study Brownian self-diffusion and structural relaxation of hard disks and compare to experimental results on quasi two-dimensional colloidal hard spheres. To this end, we calculate the self and distinct van Hove correlation functions by extending a recently proposed DDFT-approach for three-dimensional systems to two dimensions. We find that the theoretical results for both self- and distinct part of the van Hove function are in very good quantitative agreement with the experiments up to relatively high fluid packing fractions of roughly 0.60. However, at even higher densities, deviations between experiment and the theoretical approach become clearly visible. Upon increasing packing fraction, in experiments the short-time self diffusive behavior is strongly affected by hydrodynamic effects and leads to a significant decrease in the respective mean-squared displacement. In contrast, and in accordance with previous simulation studies, the present DDFT which neglects hydrodynamic effects, shows no dependence on the particle density for this quantity.

\end{abstract}

\maketitle

\section{Introduction}

Experimentally, the dynamics of colloidal suspensions in both three (3D) and two dimensions (2D) can be tracked on a single-particle level in real-time using  optical microscopy, allowing for the study of diffusive processes and structural relaxation \cite{Weeks2002, Kegel2000, Simeonova2006, Gao2007, Skinner2010, Skinner2013, Thorneywork2015_PRL, Thorneywork2016_PRL}. A key quantity in determining the temporal behavior of systems which are in a disordered phase is  the van Hove function $G(\mathbf{r}, t)$ \cite{vanHove1954, HansenMcDonald2013}, which describes the probability of finding a particle at location $\mathbf{r}$ at time $t$, given that another particle was located at the origin at $t = 0$. In particular, $G(\mathbf{r}, t)$ provides useful information to characterize phenomena such as glass- \cite{Kegel2000} or gel-transitions \cite{Gao2007}.
 
The van Hove function, or its Fourier-space counterpart, the intermediate scattering function $F(\mathbf{k}, t)$, can be obtained with some computational effort in simulations, see e.g., Refs. \cite{Puertas2004, Hopkins2010, Weysser2010,  StopperMaroltHHGRoth2015, Schindler2016, StopperHHGRoth2016JPCM}. However, in the last few years an efficient and reliable theoretical approach for calculating $G(\mathbf{r}, t)$ of a Brownian liquid has also been proposed \cite{ArcherHopkinsSchmidt2007, Bier2008} within the framework of classical density functional theory (DFT) \cite{Evans1979}, and its dynamical counter part (DDFT) \cite{MarconiTarazona1999, ArcherEvans2004}.
Here, the treatment of the important reference system of 3D hard spheres turned out to be challenging, as different models gave rise to controversial results. This included a freezing of the dynamics at unrealistically low fluid packing fractions  when compared to the formally exact predictions of computer simulations such as Brownian dynamics or kinetic Monte-Carlo methods \cite{Hopkins2010, StopperMaroltHHGRoth2015}. However, by means of considering the crossover to zero dimensions i.e., a cavity that can hold at most one particle, an excess free-energy functional has been derived within the framework of Rosenfeld's fundamental measure theory (FMT) \cite{Rosenfeld1989, Roth2010} that accurately captures the behavior of the van Hove function up to high packing fractions \cite{StopperHHGRoth2015JCP}. Moreover, the method has been successfully applied to the more complex situations of a model colloid-polymer mixture \cite{StopperHHGRoth2016JPCM}.

Recently, dynamical properties of hard disks, the 2D analogue of hard spheres, were extensively investigated in experiments considering monolayers of colloidal hard spheres \cite{Thorneywork2015_PRL, Thorneywork2016_SoftMatter, Thorneywork2017PRE}. In particular, the mean-squared displacement (MSD) and self-diffusion coefficients were measured at various fluid packing fractions $\phi = \pi R^2 \rho_b$, where $R$ is the disk-radius and $\rho_b = N/A$ the particle number density, with $N$ the number of particles per area $A$. These quantities were found to coincide quantitatively with computer simulations which do not take solvent-mediated hydrodynamic interactions (HI) into account \cite{Thorneywork2015_PRL, Thorneywork2017PRE}, implying that for this quasi-2D system hydrodynamic interactions, whilst being important at short-times, do not significantly affect the self-diffusion of particles in the long-time self-diffusion regime. Results from this colloidal system thus allow for a direct mapping between experiment and theories that do not consider HI. While these experiments have been performed in the stable liquid phase, recently dynamic properties of 2D disk-like systems have been investigated at the glass transition \cite{Vivek2017} and in amorphous solids \cite{Illing2017}.

Importantly, a FMT-type excess free-energy functional for hard disks has been derived by Roth \textit{et al.} \cite{Roth2012} and extensively benchmarked against experimental results on 2D colloidal hard spheres \cite{Thorneywork2014JCP}. However, while particle dynamics and structural relaxation have previously been addressed in simulations, see e.g., Refs. \cite{Brey2006, Sheng2012, Szamel2015_glassy2D}, a theoretical calculation of the van Hove function for hard disks and subsequent comparison to experiments is still lacking. In this work, we determine the latter where we employ the theoretical description as given in Ref. \cite{StopperHHGRoth2015JCP}, which can be applied to the 2D case straight forwardly. In particular, we compare theoretical results for the self part $G_s(\mathbf{r},t)$ of the van Hove function, describing the motion of an individual test particle, to experimental results on 2D colloidal hard spheres and find that theory and experiment are in excellent agreement at long times up to area fractions of roughly $\phi=0.60$. However, for even higher packing fractions, the self part and its respective MSD given by DDFT show some clear deviations in comparison to the experimental data.
Good quantitative agreement is also found between the distinct van Hove functions of theory and experiment at low and intermediate $\phi$, although, in accordance with the behavior of the self part, deviations are observed at higher packing fractions.

The paper is organized as follows. In Sec. \ref{SecTheory} we give an overview of the theory, including a brief introduction to dynamic test particle theory, DDFT and the FMT for hard disks. Section \ref{SecMethods} briefly describes the experimental and numerical methods employed to determine the MSD and self and distinct parts of the van Hove function. Subsequently, we present our results in Sec. \ref{SecResults}, and finally in Sec. \ref{SecConclusions} we summarize our findings.

\section{Theoretical Background} \label{SecTheory}

In order to determine the Brownian dynamics of a system in thermal equilibrium, the van Hove function $G(\mathbf{r},t)$ is an important and convenient quantity \cite{HansenMcDonald2013}. It can be written as
\begin{align} \label{EqDefvanHove}
G(\mathbf{r},t) &= \frac{1}{N} \left\langle \sum_{i,j = 1}^N\delta(\mathbf{r} + \mathbf{r}_j(0) - \mathbf{r}_i(t))\right\rangle \notag\\
&\equiv G_s(\mathbf{r},t) + G_d(\mathbf{r},t)\,,
\end{align} 
where $\langle \cdot\rangle$ denotes an average over all initial conditions of the solvent, and $\delta(\cdot)$ is the Dirac delta function. Physically, $G(\mathbf{r}, t)$ describes the probability that a particle $i$ is located at position $\mathbf{r}$ at time $t$, provided that there has been another particle $j$ at the origin $\mathbf{r} = 0$ at time $t = 0$.
Furthermore, by discriminating between the cases $i = j$ and $i \neq j$, it naturally splits into the self part $G_s(\mathbf{r},t)$, describing the motion of the (test) particle initially located at the origin, and the distinct part $G_d(\mathbf{r},t)$ accounting for the behavior of the surrounding particles. In particular, for a uniform bulk system, Eq. \eqref{EqDefvanHove} depends only on the distance $r = |\mathbf{r}|$ to the origin. 

Technically, $G(r,t)$ is a correlation function on the two-particle level, similar to the radial distribution function $g(r)$. In fact, for $t = 0$, the distinct part is exactly given by $G_d(r, 0) = \rho_b g(r)$. In the limit of very large times we have $G_d(r, t\rightarrow\infty) = \rho_b$ and $G_s(r, t\rightarrow\infty) = 0$ \cite{HansenMcDonald2013}.
 For explicit calculations, it is convenient to make use of the so-called dynamic test particle theory \cite{ArcherHopkinsSchmidt2007}, which is a dynamical extension of the well-known Percus' static test particle theory \cite{Percus1962}. It treats the system as a binary mixture of species $s$ (self) consisting only of the test particle, and species $d$ (distinct) which consists of the remaining $N-1$ particles. Hence, in thermal equilibrium at time $t = 0$, the one-body density distribution $\rho_d(r, t=0)$ around species $s$ is given by $\rho_d(r, t = 0) = \rho_b g(r)$, and $\rho_s(r, t = 0) = \delta(\mathbf{r})$. Now assume that the coordinate system is fixed in space at the original position of the tagged particle, thus for times $t > 0$, one can then follow the time evolution of the densities $\rho_s(r,t)$ and $\rho_d(r, t)$. Considering the definition of $G(r,t)$, we can identify $\rho_s(r,t) \equiv G_s(r,t)$ and $\rho_d(r,t) \equiv G_d(r,t)$. 

In particular, one can employ dynamic test particle theory within the framework of DDFT in order to describe and investigate Brownian diffusion processes. For a binary mixture consisting only of the self and distinct particles, the key equation of the DDFT that we use in this work can be written in the form of a continuity equation \cite{MarconiTarazona1999, ArcherEvans2004} (we adopt the notation of Ref. \cite{StopperHHGRoth2015JCP})
\begin{equation} \label{EqDDFT}
\frac{\partial \rho_{s/d}(\mathbf{r}, t)}{\partial t} = - \nabla \cdot \mathbf{j}_{s/d}(\mathbf{r}, t)\,,
\end{equation}
where the particle current $\mathbf{j}_{s/d}(\mathbf{r}, t)$ is assumed to be driven by chemical-potential gradients \cite{ArcherEvans2004, StopperHHGRoth2015JCP}
\begin{equation} \label{EqDefCurrent}
\mathbf{j}_{s/d}(\mathbf{r}, t) = - D_{s/d}(\mathbf{r},t) \rho_{s/d}(\mathbf{r}, t) \nabla \beta \mu_{s/d}(\mathbf{r}, t)\,,
\end{equation}
with the inverse temperature $\beta = 1/(k_B T)$.  The quantities $D_{s/d}(\mathbf{r},t)$ denote space- and time-dependent diffusivites of the particles; note that this is an adjustment to standard DDFT \cite{ArcherEvans2004} which assumes $D_{s/d}(\mathbf{r},t) = D_0$, where $D_0$ is the Stokes-Einstein single-particle diffusion coefficient (see also subsequent discussion).

The local chemical potential $\mu_{s/d}(\mathbf{r}, t)$ is obtained from the functional-derivative of the \textit{equilibrium} Helmholtz free-energy functional,
\begin{equation} \label{EqDefLocalChemicalPotential}
\mu_{s/d}(\mathbf{r}, t) = \frac{\delta F[\rho_s,\rho_d ]}{\delta \rho_{s/d}(\mathbf{r},t)}\,,
\end{equation}
which is given by
\begin{align} \label{EqHelmholtzFreeEnergy}
F[\rho_s, \rho_d] &= k_B T \sum_{l = s,d} \int \text{d}\mathbf{r}\, \rho_l(\mathbf{r}, t)\left(\ln(\lambda^2 \rho_l(\mathbf{r}, t)) - 1\right) \notag \\
&+ F_\text{ex}[\rho_s,\rho_d] +  \sum_{l =s,d}\int\text{d}\mathbf{r}\,\rho_l(\mathbf{r},t)V_\text{ext}(\mathbf{r},t)\,.
\end{align}
Here, $\lambda$ is the thermal wavelength, and $V_\text{ext}(\mathbf{r}, t)$ is an arbitrary (time-dependent) external potential. Note that in order to determine $G_d(r,0)$, one demands that $V_\text{ext}(r)$ equals the particle interactions, but vanishes for all times $t>0$. The first part of Eq. \eqref{EqHelmholtzFreeEnergy} is the exactly known ideal-gas contribution, whereas the intrinsic excess free-energy functional $F_\text{ex}[\rho_s, \rho_d]$ describes the specific particle interactions. In practice, the latter quantity generally is known exactly only in very few cases such as one-dimensional hard rods \cite{Percus1976, Vanderlick1989}, and in general approximations have to be made.
Note that employing Eqs. \eqref{EqDefCurrent}--\eqref{EqHelmholtzFreeEnergy} is equal to assuming that non-equilibrium two-body correlations are identical to their equilibrium counterparts, which obviously is exact for non-interaction particles \cite{ArcherEvans2004}--but provides an approximation to the dynamics of interacting particles. Nevertheless, DDFT successfully has been applied to various situations e.g. phase separation in a cavity \cite{Archer2005}, sedimentation in a gravitational field \cite{RoyallDzubiellaSchmidtBlaaderen2007}, the micro-rheology of colloid-polymer mixtures \cite{RauscherKrueger2007}, and also for hard disks in confinement \cite{Goddard2016}.

In this work, we describe the hard-disk interactions within the framework of FMT \cite{Rosenfeld1989, Roth2012}, i.e. the excess free-energy functional $F_\text{ex}$ is given by a volume integral in two dimensions over an energy density $\Phi$
\begin{equation}
\beta F_\text{ex}[\rho_s, \rho_d] = \int\text{d}\mathbf{r}\,\Phi(\{n_\alpha\})\,,
\end{equation}
which itself depends on a set of weighted densities $n_\alpha(\mathbf{r})$. These are defined as convolutions of the one-body density $\rho(\mathbf{r})$ with certain weight functions $\omega_\alpha(\mathbf{r})$,
\begin{equation}
n_\alpha(\mathbf{r}) = \sum_{i=s,d} \int\text{d}\mathbf{r}'\,\rho_{s/d}(\mathbf{r}')\omega_\alpha(\mathbf{r} - \mathbf{r}')\,,
\end{equation}
where the weight functions characterize the geometry of the particles, and can be of scalar-, vector- ($\overrightarrow{\cdot}$) or second-rank tensor-type ($\overleftrightarrow{\cdot}$). Note that while the self- and distinct particles are formally treated as different species, physically they are identical and hence the weight functions do not depend on the species. This restriction, of course, is straightforward to lift in order to treat multi-component mixtures. For instance, $\omega_2(\mathbf{r}) = \Theta(R-|\mathbf{r}|)$ describes the disk area, $\omega_0(\mathbf{r}) = \delta(R-r)$ characterizes the surface, $\overrightarrow{\omega_2}(\mathbf{r}) = \mathbf{r}\delta(R-r)/r$ a surface normal, and $\overleftrightarrow{\omega_2}(\mathbf{r}) = \mathbf{r}\mathbf{r}\delta(R-r)/r^2$ is the tensorial contribution. The free-energy density $\Phi$ reads \cite{Roth2012}
\begin{align} \label{Eq2DFMT}
\Phi(\{n_\alpha\}) &= - n_0 \ln(1-n_2) + \frac{1}{4\pi(1-n_2)} \notag \\
&\times \left(\frac{19}{12}n_0^2 - \frac{5}{12} \overrightarrow{n_2}\cdot\overrightarrow{n_2} - \frac{7}{6}\overleftrightarrow{n_2}\cdot\overleftrightarrow{n_2}\right)\,,
\end{align}
which has been proven to give accurate results for the radial distribution function $g(r)$ of hard-disk mixtures in comparison to experiments \cite{Thorneywork2014JCP}. Note that $\overleftrightarrow{n_2}\cdot\overleftrightarrow{n_2}$ denotes a full contraction of the tensors.

In previous theoretical studies addressing Brownian diffusion of three-dimensional colloidal hard spheres \cite{Hopkins2010, StopperMaroltHHGRoth2015, StopperHHGRoth2015JCP}, it turned out that the quality of the results compared to those from simulations crucially depends on (i) the choice of the underlying functional characterizing hard-body correlations, and (ii) the precise treatment of the self component $s$, which represents one single particle. Generally, the free-energy functionals in DFT are of grand-canonical character, and fluctuations are known to yield unphysical contributions in systems where particles are treated explicitly \cite{ReinhardtBrader2012, delasHeras2014}. 
Importantly, a method has been derived which removes possible self-interactions within the free-energy functional by considering the zero-dimensional crossover and a proper modeling of the grand potential in that situation. The main result is simply to subtract the contribution of the individual self particle from the full (mixture) excess free-energy functional \cite{StopperHHGRoth2015JCP}
\begin{equation} \label{EqDefDiffusion_DDFT}
F_\text{ex}^q[\rho_s, \rho_d] = F_\text{ex}[\rho_s, \rho_d] - F_\text{ex}[\rho_s]\,,
\end{equation} 
which is referred to as \lq quenched\rq\, functional (hence the superscript \lq q\rq\,). While the free-energy functional employed in \cite{Hopkins2010}, which is based on the relatively simple Ramakrishnan-Yousuff functional \cite{Ramakrishnan1979} implicitly obeys the form of Eq. \eqref{EqDefDiffusion_DDFT}, it predicts a freezing of the dynamics at unrealistic low packing fractions. 
Only FMT-based functionals along with Eq. \eqref{EqDefDiffusion_DDFT} so far have shown to not yield unsatisfactory results for the van Hove function in terms of structural mismatches between DDFT and simulations \cite{ StopperMaroltHHGRoth2015, StopperHHGRoth2015JCP}.

\begin{figure}[t!] 
	\centering
	\includegraphics[width = 9cm]{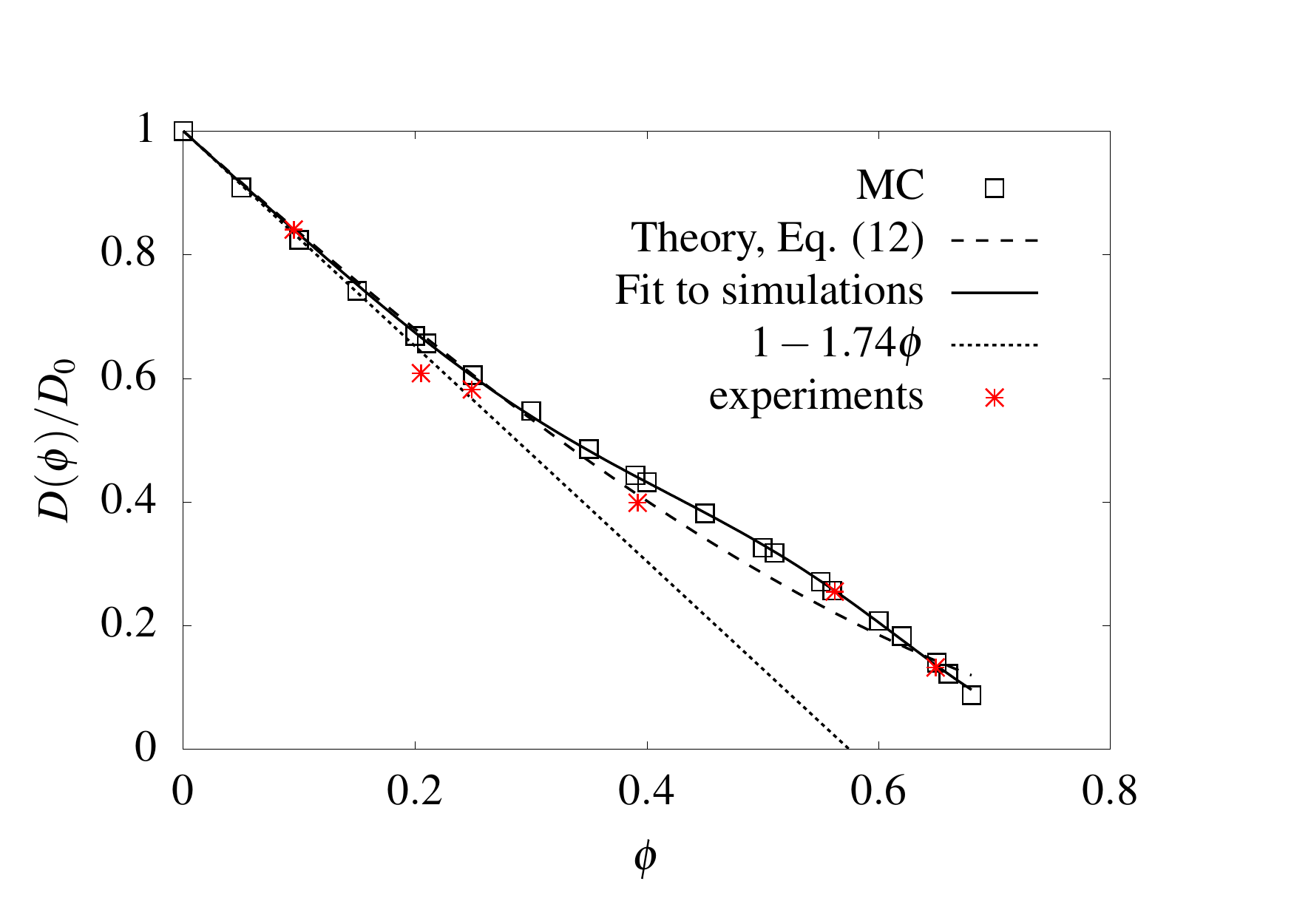} 
	\caption{Long-time self-diffusivity $D$ as a function of area fraction $\phi$ as obtained from Monte-Carlo simulations and experiments \cite{Thorneywork2015_PRL} (open squares, and red stars), the fit according to Eq. \eqref{EqFitToSim}, theoretical prediction with $\gamma = 1.68$, Eq. \eqref{EqTheoreticalD_L}, and the low packing fraction limit $1-1.74\phi + \mathcal{O}(\phi^2)$.}  \label{FigD_L}
\end{figure}

However, even with a proper removal of self-interactions, standard DDFT is not capable of adequately describing the crossover from free diffusion at very short times to the slowed down long-time self-diffusivity $D_s < D_0$. The MSD given by standard DDFT typically tends to decrease at intermediate times, but subsequently predicts the particles to speed up to exhibit ideal-gas diffusion in the long-time limit \cite{Hopkins2010, StopperMaroltHHGRoth2015, StopperHHGRoth2015JCP}. Hereby the theory predicts a behavior of the MSD which is akin to superdiffusivity, which is unphysical in the present context of overdamped equilibrium dynamics. This behaviour of the MSD is due to the fact that the decay of $G_s(r,t)$ in standard DDFT can be affected only by structural information that is encoded within the respective distinct part of the van Hove function. However, for long times, the correlations in $G_d(r,t)$ decay to a flat bulk profile and thus structural information is lost, with the \textcolor{blue}{standard} DDFT therefore yielding the (incorrect) ideal-gas diffusivity.
This shortcoming can be bypassed by assuming space- and time-dependent diffusivities $D_{s/d}(\mathbf{r},t)$ as done in Eq. \eqref{EqDefCurrent}, in order to capture effects of high packing fractions on the (self-) diffusive behavior. Note that a similar adjustment of standard DDFT for instance has also been employed in descriptions for sedimentation processes of colloidal hard spheres \cite{RoyallDzubiellaSchmidtBlaaderen2007}.
Obviously, the approach \eqref{EqDefCurrent} requires that $D(\mathbf{r}, t)$ is known at least numerically. As in a previous publication addressing diffusion in a model colloid-polymer mixture \cite{StopperHHGRoth2016JPCM}, we obtain $D(\mathbf{r}, t)$ via a suitable fit to kinetic Monte-Carlo simulations for the present system \cite{Thorneywork2015_PRL}, and subsequently by generalizing to inhomogeneous situations. The fit function that we employ is given by
\begin{equation} \label{EqFitToSim}
D(\phi)/D_0 = \exp(-1.74\phi + a\phi^3 + b\phi^4 + c\phi^5)\,,
\end{equation}
with $a = -13.25$, $b = 46.72$, and $c=-48.01$, which gives an accurate fit to the data and obeys the correct low packing fraction behavior \cite{Thorneywork2015_PRL}

\begin{equation}
	D(\phi)/D_0 = 1 - 1.74\phi + \mathcal{O}(\phi^2)\,,
\end{equation}

as is demonstrated in Fig. \ref{FigD_L}.
Note that, besides the  employed fit to simulations, one can alternatively make use of simpler theoretical approximations for $D(\phi)$. An adequate choice is a result obtained from Leegwater and Szamel that is valid in 3D and 2D \cite{LeegwaterSzamel1992}
\begin{equation} \label{EqTheoreticalD_L}
D(\phi) = \frac{D_0}{1 + \gamma\phi g(\sigma^+)}\,,
\end{equation}
where for hard disks (as well as for hard spheres) $\gamma = 2$, and $g(\sigma^+) = (1-\phi/2)/(1-\phi)^2$ is the contact value of the radial distribution function $g(r)$ from 2D scaled particle theory \cite{Helfand1961}. For instance, the authors employed Eq. \eqref{EqTheoreticalD_L} in Ref. \cite{StopperHHGRoth2015JCP} for calculating the van Hove function of 3D hard spheres, and in Ref. \cite{Thorneywork2017PRE} good agreement with present experiments was found for $\gamma = 1.68$ (see  dashed line in Fig. \ref{FigD_L}). However, here, we make use of the fit to computer simulations, which provides the most accurate theoretical description of $D(\phi)$ to be used within DDFT.

The generalization to inhomogeneous situations is achieved by replacing $\phi$ via a weighted-density approximation with $\phi\rightarrow n_2^{s/d}(\mathbf{r},t)$, where
\begin{equation} \label{EqDefn2}
n_2^{s/d}(\mathbf{r},t) = \int\text{d}\mathbf{r}\,\rho_{s/d}(\mathbf{r},t)\Theta(R - |\mathbf{r}-\mathbf{r}'|)\,,
\end{equation}
is the 2D-volume weighted density. The diffusivity of the self particle, $D_s(\mathbf{r},t) \equiv D_s(n_2^d)$, is evaluated at $n_2^d(\mathbf{r},t)$ (as it is influenced only by its surrounding neighbours when neglecting HI), whereas the distinct particles are influenced by the self- as well as surrounding other distinct particles, i.e. $D_d(\mathbf{r},t)\equiv D_d(n_2^d + n_2^s)$. As we will see in Sec. \ref{SecResults}, this weighted-density approximation yields very good results up to packing fractions of $\phi \approx 0.60$, but can gives rise to unphysical behavior at even higher densities.

\section{Experimental and Numerical Methods} \label{SecMethods}

Experimentally, the quasi-2D colloidal system was introduced in \cite{Thorneywork2014JCP}, and has previously been shown to be an excellent model for hard disks \cite{Thorneywork2014JCP,Thorneywork2015_PRL}. 
It is composed of monolayers of melamine formaldehyde particles (Microparticles GmbH), with a hard sphere diameter of $\sigma=4.04~\mu$m, that are dispersed in a 20/80 v/v$\%$ ethanol/water mixture and allowed to sediment to form a monolayer on the base of a glass sample cell. Note that the height of the sample cell is much larger than the particle diameter, $\sim50~\sigma$. The high mass density of the particles results in a gravitational length of $0.02~\mu $m and thus out of plane fluctuations are negligible and the system is structurally two-dimensional. The area fraction, $\phi$, is varied over a range from approximately $\phi = 0.1$ to 0.65. The system is imaged at a rate of 2 frames per second for up to 30 minutes using a simple video-microscopy setup, consisting of an Olympus CKX41 inverted microscope with a 40x objective and equipped with a PixeLink CMOS camera (1280 x 1024 pixels). Standard particle tracking software \cite{Crocker1996} is used to obtain particle coordinates, with an error of $12\pm10$ nm in the particle position. 

\begin{figure*}[t!] 
	\centering
	\includegraphics[width = 0.95\textwidth]{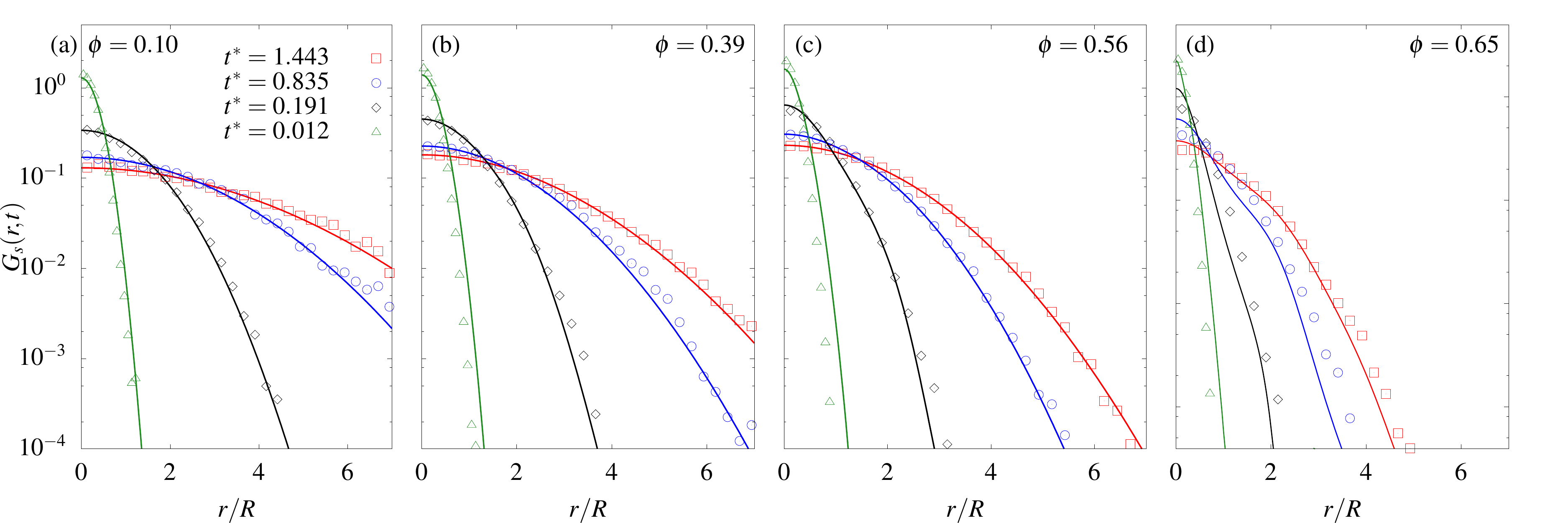} 
	\caption{Self part $\rho_s(r,t) = G_s(r,t)$ of the van Hove function for area fractions $\phi = 0.1$ (a), $0.39$ (b), $0.56$ (c) and $0.65$ (d). Individual points show experimental results at times $t^*$ = 1.443 (red squares), $0.835$ (blue circles), $0.191$ (black diamonds), and $0.012$ (green triangles). Solid lines show the predictions of DDFT. } \label{FigSelfvanHove}
\end{figure*}

The self part of the van Hove function, $G_s(x,t)$, is computed directly from particle coordinates in one spatial dimension as \cite{Thorneywork2016_SoftMatter}
\begin{equation}
\label{Eq:vanhoveexp}
G_s(x,t)= \left< \frac{1}{N} \sum^N_{j=1} 
\delta[{x}-{x}_j(t)+x_j(0)] \right>,
\end{equation}
where for a homogeneous, isotropic system in 2D, calculation of Eq.~\eqref{Eq:vanhoveexp} for movement in the $x$ or $y$ direction leads to identical results, such that $G_s(x,t)=G_s(y,t)=G_s(r,t)$. The distinct part of the van Hove function, $G_d(x,t)$, is also calculated directly from particle coordinates according to Eq.~\eqref{EqDefvanHove} with $i\ne j$. In both cases, quantities are averaged over multiple particles and time origins.

The DDFT along with the 2D-FMT is discretized on a standard rectangular 2D grid with a equidistant spacing of $\Delta = R/30$ along each axis. In order to minimize finite size effects, the number of grid points is $N = (1024)^2$ for packing fractions up to $\phi = 0.56$ and $N = (2048)^2$ for the highest packing fraction $\phi = 0.65$ . Note that while the problem is radially symmetric, which in principal allows for an analytical reduction of the FMT to effectively one dimension, the resulting equations no longer include convolutions making the numerical evaluation much more inefficient than using standard Fourier methods that can be applied in the full 2D case -- this in contrast to three dimensions, where the effectively 1D equations still include convolutions. In particular, we employ a massively in-parallel computation of the DDFT and FMT on graphics cards \cite{StopperRoth2017JCP}. 
The equilibrium density profile $G_d(r,0)$ is obtained by a standard minimization of the grand potential $\Omega[\rho] = F[\rho] + \int \text{d}\mathbf{r} \rho(\mathbf{r})(V_\text{ext}(\mathbf{r}) - \mu)$ using Percus' test particle limit, i.e. the external potential $V_\text{ext}(r)$ acting on the fluid is equal to the particle interactions. The self part $\rho_s(r,0)$ is initialized with a strongly peaked Gaussian distribution. For $t > 0$, the two-component DDFT is integrated forward in time without any external potential, $V_\text{ext}(r) = 0$. Since we are only interested in the dynamics up to times $t \sim \tau_B$, we employed a simple Euler-forward algorithm using time steps of $\Delta t = 10^{-5} \tau_B$.

\section{Results and Discussion} \label{SecResults}

Using the theory as outlined in Sec. \ref{SecTheory}, in Figs \ref{FigSelfvanHove} (a)--(d) we display the self part of the van Hove function $G_s(r,t)$ (solid lines) against the experimental results (individual symbols) for area fractions $\phi = 0.1$ (a), $0.39$ (b), $0.56$ (c), and $0.65$ (d), where the abscissae are displayed on a logarithmic scale. Note that the highest packing fraction of 0.65 is rather close to the experimentally found first-order transition from a liquid to hexatic phase at $\phi \approx 0.68$ \cite{Thorneywork2016_PRL}. Times are in reduced units, $t^* = t/\tau_B$, where $t^* = 0.012$ (green triangles), $0.191$ (black diamonds), $0.835$ (blue circles), and $1.443$ (red squares). For a low area fraction of $\phi = 0.1$, which is in the dilute limit, we see that DDFT and experiments coincide perfectly, consistent with the fact that both the DDFT itself and the excess free-energy functional become most reliable at low packing fractions. More importantly, the self part given by DDFT is also in excellent agreement with the experiments up to area fractions $\phi \approx 0.55$ for long times, as is demonstrated by Figs. \ref{FigSelfvanHove} (b) and (c). For short times, the theory is a little too broad compared to the experiments, which is to be expected, as here the decay of the experimental $G_s(r,t)$ is affected by hydrodynamic effects.
For the highest packing fraction $\phi = 0.65$, the comparison between experiment and theory shows some clear quantitative differences. In particular, at intermediate times $t^* \approx 0.2$--0.8 the DDFT predicts a clear deviation from a Gaussian shape (which would be a parabola on the logarithmic scale). While experiments also find that $G_s(r,t)$ is non-Gaussian at these times \cite{Thorneywork2016_SoftMatter}, this is not directly inferable on the scales shown here. For longer times, the agreement between theory and experiment becomes quantitatively good again.

 \textcolor{red}{}

\begin{figure}[t!] 
\centering
 \includegraphics[width = 9cm]{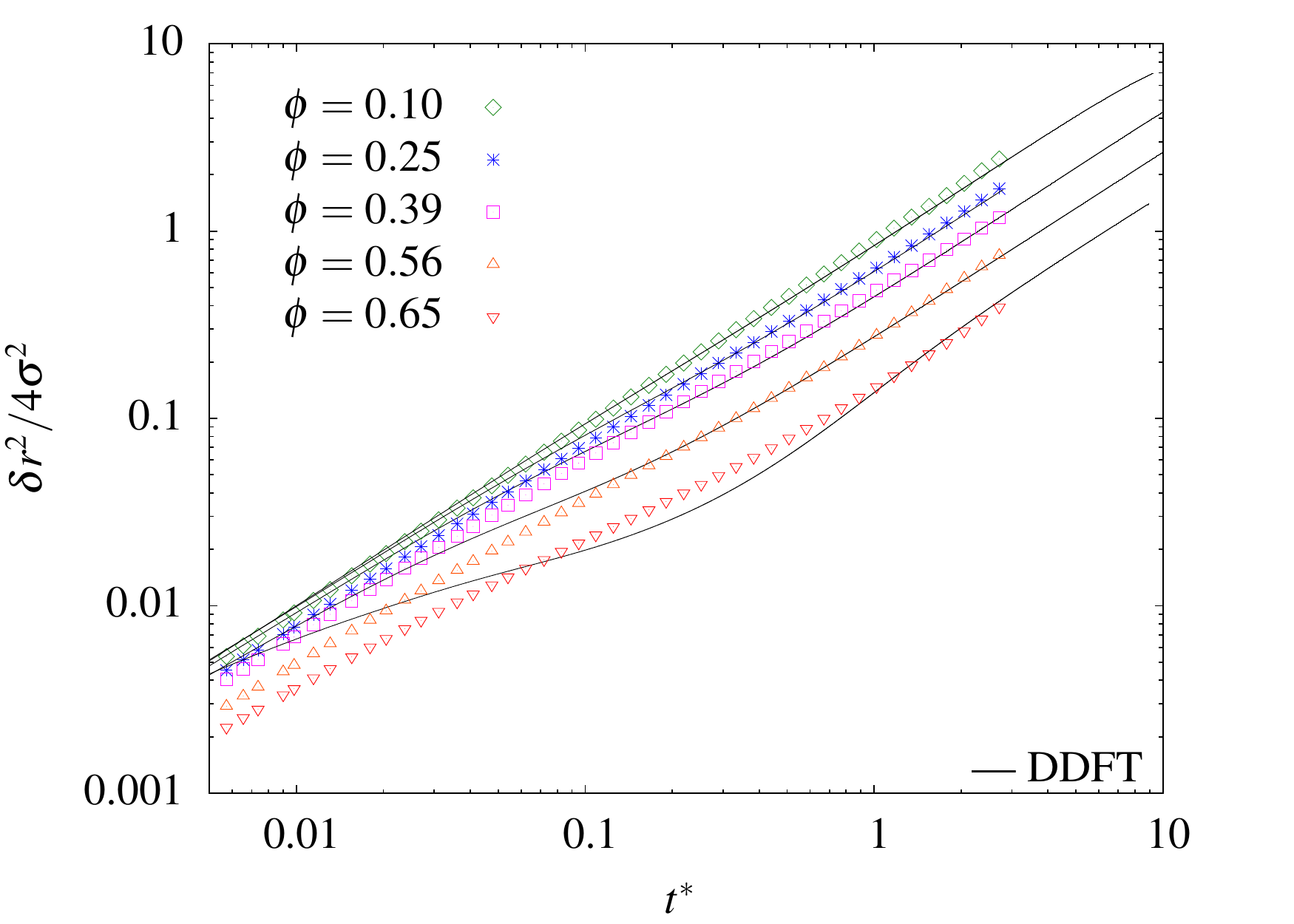} 
 \caption{Mean-square displacement $\delta r^2$ normalized to the ideal-gas versus time $t^*$ on a double-logarithmic scale of the experiments (individual symbols) for area fractions $\phi = 0.1$ (green diamond), 0.25 (blue stars), $0.39$ (pink squares), $0.56$ (orange up-triangles), and $0.65$ (red down-triangles). The black solid lines are the respective DDFT results.} \label{FigMSD}
\end{figure}

\begin{figure*}[t!] 
	\centering
	\includegraphics[width = 1.0\textwidth]{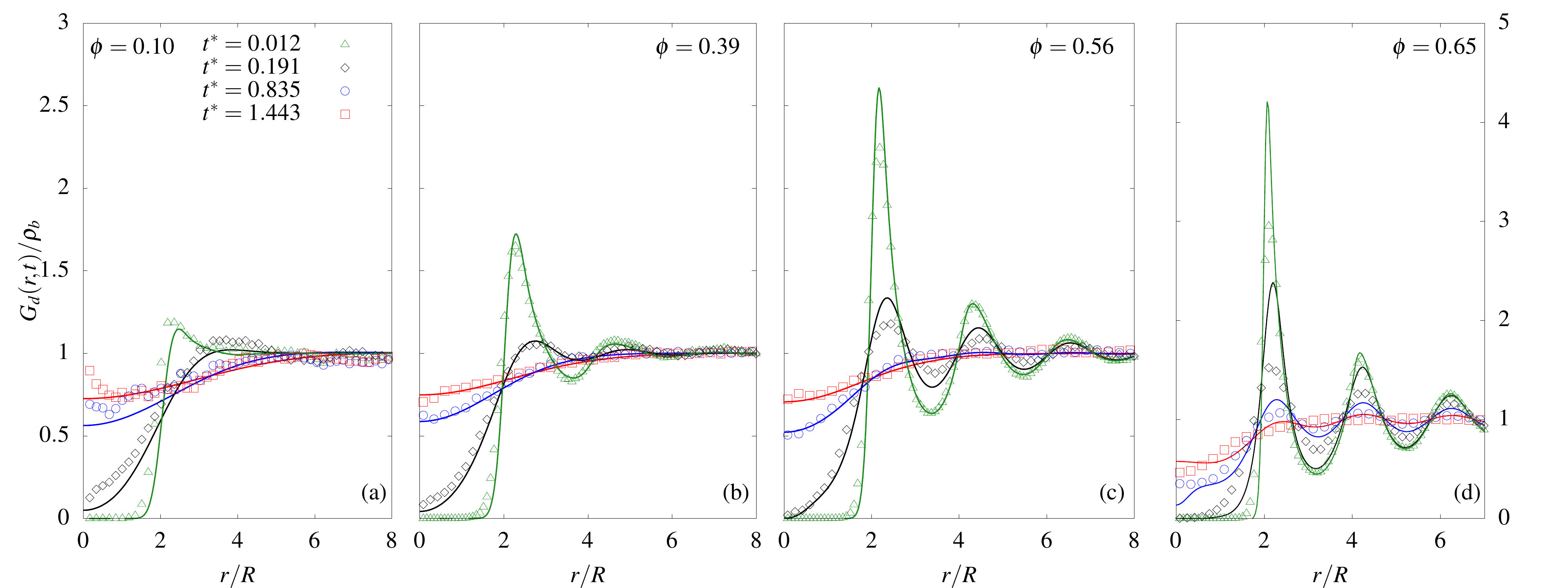} 
	\caption{Distinct part $\rho_d(r,t) = G_d(r,t)$ of the van Hove function for area fractions $\phi = 0.1$ (a), $0.39$ (b) $0.56$ (c), and $0.65$ (d) as borne out by the present DDFT at times $t^* = 0.012$ (green), 0.191 (black), 0.835 (blue) and 1.443 (red) in comparison to experimental results (symbols). } \label{FigDistinctvanHove}
\end{figure*}

In Fig. \ref{FigMSD} we plot the respective MSD, $\delta r^2(t)$, normalized w.r.t the ideal-gas result $\delta r_\text{ideal}^2(t) = 16R^2t^*$. The MSD is obtained from $G_s(r,t)$ (solid lines) via
\begin{equation} \label{EqDefMSD}
\delta r^2(t) = \int \text{d}\mathbf{r}\,\mathbf{r}^2 G_s(r,t)\,,
\end{equation}
whereas experimentally the MSD is obtained from its statistical definition $\delta r_\text{exp}^2(t) = \langle[\mathbf{r}(t)-\mathbf{r}(0)]^2\rangle$, where $\mathbf{r}(t)$ is the position of a tagged test-particle at time $t$ and $\langle\cdot\rangle$ is an average over distinct time origins and all particles.
The experimental area fractions are $\phi = 0.1$ (green diamonds), $0.25$ (blue stars), $0.39$ (pink squares), $0.56$ (orange up-triangles), and $0.65$ (red down-triangles). Note that the renormalization with respect to the infinite-dilution single-particle diffusion coefficient is a deliberate choice: in Ref. \cite{Thorneywork2015_PRL} it has been demonstrated that at long times, the reduction in the diffusion coefficient with increasing density is precisely that expected from only the effect of direct interactions i.e. that the reduction in the short-time diffusion coefficient due to solvent-mediated hydrodynamic interactions (HI) does not change the time scales relevant at long times. Hence, to compare experiments to theory (or simulations) it is only necessary to take into account the single particle diffusivity.

As indicated by the self part of the van Hove function, $G_s(r,t)$, we see that theory accurately captures the experimental long-time behavior up to $\phi = 0.56$. In contrast, the short-time dynamics in the experiments are clearly slowed down with respect to the DDFT, due to the fact that with increasing $\phi$, experimentally the short-time relaxation is found to be significantly affected by HI (see Ref. \cite{Thorneywork2015_PRL}). Physically, this corresponds to the regime where the test (or, equivalently, self particle) has not experienced direct interactions with the shell of neighbouring particles but has been effected by HI. The agreement between the MSD from experiment and theory at long times but not at short times again highlights that HI effectively do not effect the long time self-diffusion in the experiment i.e. that the reduction in the long-time diffusion coefficient of the particles can be accounted for by direct interactions alone, and does not appear to reflect the reduced short time mobility of the particle. 

The differing short-time behavior between DDFT and experiments results in a significant crossover from free diffusion to the slowed down long-time self-diffusion at intermediate times $t^* \approx 0.05$--0.5 in the DDFT results, which is less pronounced in the experiments. We have verified that the behavior of the MSD given by the theory is in good quantitative agreement with simulations up to $\phi \approx 0.56$ for all times (not shown here). However, as indicated by the behavior of the self part, at $\phi = 0.65$ the respective MSD seems to capture the desired long-time diffusivity, but its intermediate time behavior between $t^* \approx 0.3$ and 1.0 shows a curve with a slope that is greater than unity. Considering that the present DDFT aims to describe overdamped equilibrium dynamics, such a behavior of the MSD is strictly speaking unphysical. This inadequacy can be traced back to the weighted-density approximation employed in order to obtain a space- and time-dependent diffusivity $D_{s}(r,t)$: for $\phi = 0.65$, the density distribution of $G_d(r,t)$ exhibits strong oscillations (cf. Fig. \ref{FigDistinctvanHove} (d))  at times $t^* \approx 0.1-0.5$ resulting in a highly inhomogeneous $D_{s}(r,t)$, thereby overestimating the slowdown of the corresponding MSD. We are confident that the observed deviations of $G_s(r,t)$ from a Gaussian shape are, at least by parts, related to this shortcoming as standard DDFT with $D_{s/d}(r,t) = D_0$ predicts much less pronounced structural deviations. Note that the short-time behavior of $G_s(r,t)$ is only weakly affected by this defect as $G_d(r,t) \approx 0$ in vicinity to the origin for $t\rightarrow 0$. 	
For longer times, however, when the correlations have sufficiently flattened out, the diffusivity increases again. As a result, the MSD finally approaches the prescribed long-time diffusion thereby predicting \lq superdiffusivity\rq\,. From a technical point of view, this behavior is akin to the incapability of standard DDFT capturing the correct long-time asymptotics (cf. discussion in Sec. \ref{SecTheory}). Note that one obtains a similar behavior when e.g. making use of the theoretical expression Eq. \eqref{EqTheoreticalD_L}.

In Figs. \ref{FigDistinctvanHove} (a)--(d) we show the distinct part of the van Hove function $G_d(r,t)$ for the same area fractions as in Figs. \ref{FigSelfvanHove} (a)--(d) and times [$t^*=0.012$ (green), 0.191 (black), 0.835 (blue), and 1.443 (red)]. Again, we find excellent agreement between the experimental data and the DDFT for all times for low and intermediate packing fractions (cf. Figs. \ref{FigDistinctvanHove} (a) and (b)). Note that the pronounced fluctuations in the experimental data for $\phi=0.1$ are due to poorer statistics arising from the smaller number of particles at this packing fraction compared to higher packing fractions. For a higher packing fraction, $\phi=0.56$, the agreement is satisfactory in the short- as well as in the long-time limit. However, at intermediate times, the distinct part given by the DDFT (slightly) overestimates the  particle-correlation peaks. This phenomenon has also been observed using the present theoretical framework in dense 3D hard-sphere systems \cite{StopperHHGRoth2015JCP}.
For $\phi = 0.65$, the theory predicts a significantly slower relaxation of the distinct part at intermediate times $t^* \approx 0.1$--$0.3$ compared to the experiments--see Fig. \ref{FigDistinctvanHove} (d)--in accordance with the behavior of the self part and its corresponding MSD. Similarly, for long-times, the deviations between theory and experiment again become smaller, which is in line with  the (unphysical) speed-up of the MSD to the correct long-time diffusivity.
	
Interestingly, the agreement between theoretical and experimental results for the distinct part $G_d(r,t)$ is quantitatively satisfactory for all area fractions at short times, $t^* \approx 0.01$, implying that for this quantity there is no systematic effect from hydrodynamic interactions. This is in contrast to the behavior of the self part $G_s(r,t)$ at short times. 
In Ref. \cite{BleibelOettel2015} DDFT and Stokesian dynamics simulations were used to investigate the effect of HI on collective diffusion in a system where colloidal particles are confined to a plane, but solvent flows in full 3D. Here, it was found that the density profiles exhibit an algebraic decay at large distances. However, if we plot our experimental data as $\ln(rh(r,t))$, where $h(r,t)=G_d(r,t)-1$,  we do not observe a systematic deviation from the usual exponential decay of the correlations. This may arise for several reasons. First, in Ref. \cite{BleibelOettel2015} the interactions between the particles have been neglected, which is completely inadequate for high packing fractions. Secondly, in \cite{BleibelOettel2015} the solvent flows in full 3D, i.e. here the Oseen tensor can be applied. However, in our experiments the solvent flow is much more complicated due to the impenetrable plane in which the colloids are settled on. A theoretical model including HI for the present situation thus should make use of the Blake tensor \cite{Blake1971} instead of the Oseen tensor, which probably would lead to a reduction of the impact of hydrodynamic effects, as is demonstrated e.g. in Ref. \cite{Goddard2016}. 
 Thus, elucidating the possible effects of HI on collective quantities such as $G_d(r,t)$ for the present system would require a much more careful and detailed analysis of both theory and experiments, which goes beyond the scope of the present work. 
Considering the overall behavior of the van Hove function $G(r,t)$, we find that DDFT in its present (nearly standard) form along with the quenched functional \eqref{EqDefDiffusion_DDFT}  performs remarkably well in describing bulk dynamics of colloidal hard disks compared to experiments, provided that the long-time diffusivity is adequately accounted for as an input and the packing fractions become not too high. We note that, in contrast, the recently developed framework of power functional theory (see, e.g., Refs. \cite{SchmidtBrader2013, BraderSchmidt2014, SchindlerSchmidt2016}) is in principle exact and should yield the correct long-time dynamics without employing empirical inputs. This is because, unlike DDFT, it accounts for memory of the past motion of particles. However, up to this date, power functional theory seems not to be of practical use for dynamic test particle calculations as explicit expressions for the power functionals are still currently under development \cite{deLasHerasSchmidt2018}.

\section{Summary and Outlook} \label{SecConclusions}

We studied the time evolution of the self and distinct parts of the van Hove function for Brownian hard disks by means of dynamical density functional theory within the framework of dynamic test particle theory. To this end, we employed an accurate FMT-type excess free-energy functional \cite{Roth2012, Thorneywork2014JCP} along with the so-called \lq quenched\rq\, approach \cite{StopperHHGRoth2015JCP}, which has proven to yield accurate results in dense 3D hard-sphere systems. 

The results of the theory for the self part of the van Hove function, $G_s(r,t)$, and the corresponding mean-squared displacement are directly compared to experimental data on 2D colloidal hard spheres (Figs.~\ref{FigSelfvanHove} and \ref{FigMSD}), where we find excellent quantitative agreement up to relatively high fluid packing fractions of roughly 0.60 at long times. Up to this density regime, we also find good quantitative agreement between experiment and theory for the distinct van Hove function at all times. At very high fluid packing fractions, close to the experimentally found liquid-hexatic transition, the present DDFT becomes less reliable, manifested in clear deviations of the self part from a Gaussian  shape or unphysical behavior of the MSD at intermediate times. However, it is important to note that the observed deviations at high volume fractions are not to be interpreted as a precursor of nucleation or crystallization, but merely are an artifact of the theoretical approach. In particular, the input of the fitted long-time diffusivity along with the empirical adoption to the inhomogeneous fluid seems to become rapidly inadequate for packing fractions $\phi \gtrsim 0.60$.

In the short-time self-diffusion regime, which in the experiments is strongly affected by solvent-mediated hydrodynamic interactions \cite{Thorneywork2015_PRL}, deviations are found between the experiment and DDFT, as the latter predicts ideal-gas diffusion and does not explicitly take HI into account. While the framework of DDFT in principle allows for the treatment of HI, see, e.g., Refs. \cite{RexLoewen2009} for concrete applications, in the present case it is a highly non-trivial task to theoretically model the boundary conditions exerted by the experimental geometry on the solvent flow -- although a possible approach has recently been presented in Ref. \cite{Goddard2016}. Nevertheless, from a theoretical point of view, our present results confirm that dynamic test particle theory along with DDFT and a proper removal of self interactions according to Eq. \eqref{EqDefDiffusion_DDFT}, form a valuable and adequate tool for investigating the structural relaxation of hard disks subject to Brownian motion in the overdamped limit. 
Future interesting questions, beside modelling HI for the present experimental situation properly, include, e.g., the dynamics in confined geometries or an incorporation of additional long-ranged forces.

\begin{acknowledgments}
Funding by the Carl-Zeiss-Stiftung, the EPSRC and the ERC (ERC Starting Grant No. 279541-IMCOLMAT) is gratefully acknowledged. 
\end{acknowledgments}

\bibliographystyle{apsrev}
\bibliography{references_2}

\end{document}